\documentclass[preprint]{aastex}
\begin{document}

\font\twelvei = cmmi10 scaled\magstep1 
       \font\teni = cmmi10 \font\seveni = cmmi7
\font\mbf = cmmib10 scaled\magstep1
       \font\mbfs = cmmib10 \font\mbfss = cmmib10 scaled 833
\font\msybf = cmbsy10 scaled\magstep1
       \font\msybfs = cmbsy10 \font\msybfss = cmbsy10 scaled 833
\textfont1 = \twelvei
       \scriptfont1 = \twelvei \scriptscriptfont1 = \teni
       \def\mit{\fam1 }
\textfont9 = \mbf
       \scriptfont9 = \mbfs \scriptscriptfont9 = \mbfss
       \def\bmit{\fam9 }
\textfont10 = \msybf
       \scriptfont10 = \msybfs \scriptscriptfont10 = \msybfss
       \def\bmsy{\fam10 }

\def\etal{{\it et al.~}}
\def\eg{{\it e.g.,~}}
\def\ie{{\it i.e.,~}}
\def\lsim{\raise0.3ex\hbox{$<$}\kern-0.75em{\lower0.65ex\hbox{$\sim$}}} 
\def\gsim{\raise0.3ex\hbox{$>$}\kern-0.75em{\lower0.65ex\hbox{$\sim$}}} 
\def\kms{~{\rm km~s^{-1}}}
\def\cm3{~{\rm cm^{-3}}}
\def\yr{~{\rm yr}}
\def\Msun{~{\rm M}_{\sun}}

% My def.
\def\cf{{\it cf.~}}
\def\cc{CCs~}
\def\coll{ \tau_c}
\def\tma{ \tau_{ma}}
\def\tdr{ \tau_{dr}}
\def\tcr{ \tau_{cr}}
\def\tcs{ \tau_{cs}}
\def\ssp{c_{si}}
\def\roi{ \rho_i}
\def\ycm{ $Y_{cm}$~}
\def\pma{ \p_{max}}
\def\xcoor{{\it x}-coordinate~}
\def\ycoor{{\it y}-coordinate~}
\def\zcoor{{\it z}-coordinate~}
\def\ltsima{$\; \buildrel < \over \sim \;$}
\def\simlt{\lower.5ex\hbox{\ltsima}}
\def\gtsima{$\; \buildrel > \over \sim \;$}
\def\simgt{\lower.5ex\hbox{\gtsima}}

\title{Enhanced Cloud Disruption by Magnetic Field Interaction\altaffilmark{1}}

\author{G. Gregori,\altaffilmark{2,3}
	Francesco Miniati,\altaffilmark{3}
        Dongsu Ryu,\altaffilmark{4}
and     T.W. Jones\altaffilmark{3}}

\altaffiltext{1}{Animations and color images from this work have been posted
at URL http://www.msi.umn.edu/Projects/twj/mhd3d/}

\altaffiltext{2}{Department of Mechanical Engineering, University of Minnesota,
    Minneapolis, MN 55455; gregori@me.umn.edu.}
\altaffiltext{3}{School of Physics and Astronomy, University of Minnesota,
    Minneapolis, MN 55455; min@msi.umn.edu, twj@astro.spa.umn.edu.}
\altaffiltext{4}{Department of Astronomy \& Space Science, Chungnam National
University, Daejeon 305-764, Korea; ryu@canopus.chungnam.ac.kr.}

\begin{abstract}

We present results from the first 
three-dimensional numerical simulations of moderately supersonic cloud motion
through a tenuous, magnetized medium. We show that the interaction of the
cloud with a magnetic field perpendicular to its motion 
has a great dynamical impact on the
development of instabilities at the cloud surface. Even for initially
spherical clouds, magnetic field lines become trapped in surface
deformations and undergo stretching. The consequent field amplification 
that occurs there 
and particularly its variation across the cloud face
then dramatically enhance the growth rate
of Rayleigh-Taylor unstable modes, hastening the cloud disruption.

\end{abstract}


%\clearpage

\section{Introduction}
This letter focuses on the results of the first three-dimensional (3-D) study 
of the ballistic interaction of a moderately supersonic dense cloud with a 
warmer {\it magnetized} tenuous medium. 
Magnetic fields are ubiquitous in astrophysical environments and 
cannot be neglected in any realistic ISM related study.
Mac Low \etal (1994) and Jones \etal (1996) have 
studied the two-dimensional (2-D) magnetohydrodynamics (MHD) of
cosmic bullets, extending earlier gasdynamical simulations and addressing the
importance of magnetic field in preventing the cloud disruption due to
fluid instabilities. In particular Jones \etal (1996) found that
a region of strong magnetic pressure (or magnetic bumper) develops
at the nose of the cloud during its motion 
through a magnetized medium, whenever
its velocity is perpendicular to the initial unperturbed background field
(transverse case). This
enhancement was attributed to a strong stretching of the magnetic field lines. 
Miniati \etal (1999b) further investigated the problem pointing out
that the behavior observed in the transverse case was typical, except in the
case of a very small angle ($< 30 ^o$) between the cloud velocity and the 
initial magnetic field.  In addition, the development of this magnetic shield 
in front of the cloud was recognized to play a crucial role in the outcome
of cloud collisions (Miniati \etal 1999a), because it dramatically
reduced the degree of cloud disruption. 
Further, it was important in terms of
exchange of magnetic and kinetic energy
between different phases in the ISM (Miniati \etal 1999b) and  to
highlight cosmic ray electrons (Jones \etal 1994).

While supersonic motion of an individual cloud is obviously idealized,
there are numerous astronomical contexts which can be illuminated
by its example. For example,
MHD wind-cloud interaction may be important for
the evolution of synchrotron emission processes from planetary
nebulae (Dgani \& Soker 1998).
Recently a multi-phase structure has been proposed
for the dynamical state of the cosmic intra cluster medium (Fabian 1997), 
in order to explain EUV radiation in excess to what expected from the hot 
gas there (\eg Bowyer, Lieu \& Mittaz 1998). Although the issue has not
been
settled yet, this suggestion reinforces the necessity to understand the proper
evolution of clouds in a multi-phase medium.

All the previous MHD  cloud-motion results were based 
on 2-D numerical simulations, so
the importance of following them up with more realistic 3-D
calculations is clear. Stone and collaborators (Stone \& Norman 1992;
Xu \& Stone 1995) have reported 3-D hydrodynamical simulations of
shocked gas clouds. They found qualitative agreement with analogous 2-D
gasdynamical simulations. 
However, as we shall see, introduction of nonisotropic Maxwell stresses
produces very significant effects that differ in 3-D from what has been seen 
in previous, 2-D MHD simulations.

%%\centerline{\null}
%%\vskip2.85truein
%%\special{psfile=fig1.ps angle=0 voffset=-215 hoffset=-140
%%hscale=80 vscale=80}
%%\figcaption{Volume rendering (log scale) of the magnetic pressure
%%for the $\beta=4$ simulation (strong field) at $t=2.25 \tau_{cr}$.
%%Superimposed in the figure are the magnetic field lines.
%%\label{f_lines}}

%%\vskip0.2truein

%%\begin{figure}
%%\vspace{-0.3truein}
%%\epsfysize=4.5in
%%\epsfbox[0 0 200 550]{fig1.ps}
%%\vspace{-2.1truein}
%%\caption[]{\label{f_lines}
%%Volume rendering (log scale) of the magnetic pressure
%%for the $\beta=4$ simulation (strong field) at $t=2.25 \tau_{cr}$.
%%Superimposed in the figure are the magnetic field lines.}
%%\end{figure}

\section{Numerics \& Problem Setup}
\label{code}
The numerical computation is based on a total variation diminishing
(TVD) scheme for ideal MHD (Ryu \& Jones 1995; Ryu, Jones \& Frank 1995;
Ryu \etal 1998). 
The cloud, initially spherical,
is set in motion with respect to the uniform ambient medium with a
velocity aligned along the $x$-axis. Its velocity is $u_c = M c_s$, where
$c_s$ is the sound speed in the ambient medium and the Mach number is
$M=1.5$. The initial cloud density is $\rho_c = \chi \rho_i$, with
$\rho_i$ the intercloud density and $\chi = 100$.
The direction of the magnetic field is chosen along the $y$-axis. 
Its intensity is conveniently expressed in terms of the familiar parameter
$\beta = p/p_B$, where $p$ is the hydrodynamic pressure 
and $p_B=B^2/(8 \pi)$ is
the magnetic pressure. 
In these numerical simulations we have considered both the cases of a
strong field ($\beta=4$) and a weak field ($\beta=100$).
To be able to
compare with pure hydrodynamic effects, a case with $\beta=\infty$
(no magnetic field) has also been computed.
The calculations have been performed using a $416\times 208\times 416$
zone box containing
one quarter of the physical space with radial
symmetry applied in the $y$-$z$ plane around the $x$ axis.
The initial cloud radius spanned 26 zones.
Additional details on the numerical setup can be found in a companion
paper describing quantitative results (Gregori \etal 1999).
Since the cloud motion is supersonic, its motion leads to the formation of
a forward, bow shock and a reverse, crushing shock propagating through 
the cloud. The approximate time for
the latter to cross the cloud is referred to as the 
``crushing time''\footnote{
This form of the crushing time differs from the one of Klein \etal
(1994) by a factor of 2 since our definition is based on the cloud
diameter instead of the cloud radius. We use this definition since it
more closely measures the actual time before the crushing shock emerges.}
(\eg Jones \etal 1994):
$
\tau_{cr} = 2 R_c \chi^{1/2}/M c_s
$,
where $R_c$ is the initial cloud radius. As we will see, the crushing time
is a relevant dynamical quantity since it is proportional to the 
time over which disruptive instabilities develop.

\section{Cloud Disruption}
\label{cd}
As the cloud moves through the ISM, its surface
is subjected to several instability mechanisms:
namely, Kelvin-Helmholtz (K-H), 
Rayleigh-Taylor (R-T) and, at start-up, the Richtmyer-Meshkov (R-M)
instability. 
These instabilities, but especially the R-T instability, will ultimately 
disrupt the entire cloud. 
Grid-scale noise provides the seed for initial development of
such instabilities
(Kane \etal 1999 and references therein). At the same time, 
finite numerical resolution may suppress 
the large wavenumber (small scale) mode
components that have the fastest initial growth rate (Chandrasekhar 1961).
The K-H instability is related with the shearing motion at the separation
between two fluids. Typically, we should expect such an instability
to develop on the lateral boundaries of the cloud. From our simulations we
can see that its effects are generally limited, 
especially considering that, in the MHD case,
the development of the instability tends to be suppressed by the
magnetic field for wave vectors in the $x$-$y$ plane (Chandrasekhar 1961;
Jones \etal 1996).
The most disruptive instability for a supersonic cloud is 
the R-T. It develops at the interface between two fluids when the
lighter accelerates the heavier one.
In the limiting case of an impulsive acceleration,
as at the beginning of our simulations where the clouds are set into
instantaneous motion,
the R-M instability applies. However, the linear growth of the R-M
instability and the presence of a thin boundary layer on the initial
cloud makes the R-M instability relatively unimportant.
Before a time $\lesssim \tau_{cr}$ the cloud interface
remains R-T stable, since there is no ongoing acceleration
of the cloud body to drive this instability. 
When the crushing shock exits the cloud, however,
a pressure gradient between the front and the rear is established. 
The cloud body is then decelerated, 
and this induces the R-T growth at its front
interface (\eg Mac Low \etal 1994; Jones \etal 1994; Kane \etal
1999).

In the general hydrodynamic 
case a R-T perturbation of wavenumber $k$ grows with a characteristic time
$
\tau_{RT} \simeq (g k)^{-1/2}
$,
where $g$ is the cloud deceleration (Chandrasekhar 1961).
Following Klein \etal (1994), we can estimate the deceleration as
\begin{equation} \label{decel}
g \simeq \frac{\rho_i u_c^2}{\rho_c R_c},
\end{equation}
with the cloud velocity $u_c$ and the intercloud medium assumed at
rest.
It is easy to show that the most destructive modes, those with the 
largest wavelength, $k\sim 2\pi/R_c$, develop on a timescale 
$\tau_{RT} \sim \tau_{cr}$. Thus, in the absence of magnetic field
influences, the cloud will be disrupted only on a time scale of
several $\tau_{cr}$ (Jones \etal 1996; Xu \& Stone 1995). 

%%\begin{figure*}
%%\vspace{-0.3truein}
%%\epsfysize=9.5in\epsfbox[-20 0 530 795]{fig2.ps}
%%\vspace{-0.1truein}
%%\caption[]{\label{den_hb}
%%Volume rendering of the cloud density (log scale) for the
%%$\beta=\infty$ simulation (left column), $\beta=100$ (center column), and
%%$\beta=4$ (right column). 
%%Time is expressed in units of $\tau_{cr}$.}
%%\end{figure*}

\section{Results}
\label{res}
It is usually pointed out in the literature that in the interaction of a
cloud-like object with a magnetized wind, the stretching of magnetic
field lines draping around the body of the cloud is limited by their ability to
escape by slipping around it (\eg Mac Low \etal 1994; Dgani \& Soker 1998).
On the contrary, our results show that the magnetic field influences
the development of the cloud deformations produced
by instabilities so that the field lines become trapped in such deformations.
This trapping of the field lines is indeed the most important physical
process that has been revealed from our 3-D simulations. Its main consequence
is then the development of a region of strong magnetic pressure at the
leading
edge of the cloud. 
This is clearly visible in Fig. \ref{f_lines},
which shows that even at simulation end most of the ``overrun'' field lines
are indeed kept within the R-T fingers without slipping away. 

This behavior is different from what is seen at the earth's magnetopause
or as predicted for the magnetopause of comets (\eg Yi \etal 1996).
There, field lines do seem to slip past the object, so that it behaves
somewhat like a rigid body. Indeed, for both of those cases the impacted
body is stiff, because compression produces an increasingly greater
restoring force. 

In Fig. \ref{den_hb} we compare the cloud evolution for the three
studied
cases of the field intensity through 3-D volume rendering images of gas
density. The four pictures in
the (first, second, third) columns correspond to different evolutionary times
for the progression of cases: $\beta =(\infty, 100, 4)$ respectively.
The most significant conclusion is that the 
presence of a strong background magnetic
field ($\beta=4$)
dramatically modifies the dynamical evolution of the cloud.
For the hydrodynamic simulation the cloud is initially uniformly 
crushed in the direction aligned with its motion and the
disruptive instabilities, which are radially symmetric because of
the imposed symmetry, become evident only by the end of the
simulation ($t = 2.25 \tau_{cr}$). Our resolution is not sufficient
to allow really fine K-H instabilities to develop on the cloud perimeter
such as those evident in the hydrodynamical shocked-cloud results of 
Xu \& Stone (1995), but the
global evolutions in our hydrodynamic simulations are comparable.
By contrast with this behavior, the presence 
of a strong field 
($\beta=4$ case)  causes the cloud 
to be additionally compressed along the $y$-axis by the draping magnetic
field lines. 
As a result the cloud is ``extruded'' in a direction orthogonal to
the plane containing its motion and the initial field.
This behavior is already well visible after one crushing time. 

As noted earlier, magnetic fields inhibit the R-T instabilities in the 
$x$-$y$ plane (Jones \etal 1996). 
The development of this instability only along the $z$-axis, then produces the
C-shaped structure of the cloud (last panel in Fig. \ref{den_hb}), as opposed
to the symmetrical ``sombrero'' surface of the hydrodynamical case.
In addition, the presence of the magnetic field enhances the growth of the R-T
instability. In the strong field case ($\beta=4$), this can be assessed by
replacing the ram pressure $\rho_i u_c^2$ with an effective pressure
$\rho_i u_c^2 + p_B$ in eq. (\ref{decel}). 
This is motivated by the fact that, for small $\beta$, $p_B$ becomes 
comparable to or larger than the ram pressure $\rho_i u_c^2$ during times 
$t \lesssim \tau_{cr} \sim \tau_{RT}$ (see also Miniati \etal 1999b).
Taking into account this additional magnetic term 
then shortens the time scale for the development of 
the instability by a factor $\sim 2$. The instability enters the nonlinear
stage sooner and its exponential growth determines the dramatic differences
from the hydrodynamical case as shown in Fig. \ref{den_hb}.
In this sense, our results describe the cloud disruption by a
{\it magnetically enhanced R-T like instability}. 

A weak field ($\beta=100$) changes
the cloud dynamics in a similar way, but more slowly and less dramatically.
With a limited growth of a C-shaped structures,
the development of the R-T instability in front of the cloud looks
typical, although it still shows evidence of the asymmetric Maxwell
stresses.
Also at $t\simeq \tau_{cr}$ the magnetic pressure is still a
small fraction of the ram pressure in front of the cloud. Therefore, although
the shape of the cloud undergoes some deformation compared to the
hydrodynamical case, the growth rate of the instability has not been enhanced
significantly. 
Further discussions on the magnetic energy
evolution and its effect on the cloud
morphology are given by Gregori \etal (1999).

\section{Summary}
\label{cs}

In this paper we have presented the first results of a series
of 3-D MHD numerical simulations of cloud motion in a multi-phase
interstellar medium. We have considered a spherical cloud
that moves transverse to the magnetic field, with two different
cases for its initial strength; namely $\beta = 4$ and $\beta = 100$.

Both the weak ($\beta = 100$) and strong field ($\beta = 4$) 
simulations showed a comparable behavior with
a substantial enhancement of the magnetic pressure at the leading edge
of the cloud as a result of field line stretching there. 
This confirms and extends previous 2-D results
(Jones \etal 1996; Miniati \etal 1999b).
The importance of field line stretching to field
amplification and cloud motion can now be fully appreciated, since
the slipping of the field lines around the cloud
(which was prevented in 2-D simulations by their geometry)
turned out to be a minor effect in our 3-D simulations. In general moving
clouds are reshaped by the field
lines draping around their bodies, generating elongated structures oriented
perpendicular to both the background magnetic field and the flow velocity. 
This is also important in terms of cloud collisions
because it means that if two clouds are approaching each other moving through
the same large scale magnetic field, their elongated, cylindrical
bodies will tend to collide with the longer axis aligned. This helps
validate previous 2-D
MHD cloud collision calculations (Miniati \etal 1999a) and sets important
constraints for future work on that subject.

{\it The main result of this study is that contrary to 2-D geometry
where fluid instabilities were prevented by the growth 
of a strong magnetic field, 
in 3-D they are instead considerably hastened by it.} This was clearly shown 
in Fig. \ref{den_hb} where
the dramatic difference between the purely hydrodynamic and the $\beta=4$ MHD 
case is attributed to the accelerated development of a R-T
instability in the latter.
It was also shown that for $\beta =4$ the timescale for
the cloud disruption is reduced by a factor $\sim 2$.

\acknowledgments
This work is supported at the University of Minnesota by the NSF through
grants AST 96-19438 and INT95-11654, by NASA grant NAG5-5055, and by the
Minnesota Supercomputing
Institute. Work by DR is supported  in part by KOSEF through grant
981-0203-011-2.

\begin{center}
{\bf FIGURE CAPTIONS}
\end{center}

\figcaption[]{Volume rendering (log scale) of the magnetic pressure
for the $\beta=4$ simulation (strong field) at $t=2.25 \tau_{cr}$.
Superimposed in the figure are the magnetic field lines.
\label{f_lines}}

\figcaption[]{\label{den_hb}
Volume rendering of the cloud density (log scale) for the
$\beta=\infty$ simulation (left column), $\beta=100$ (center column), and
$\beta=4$ (right column). 
Time is expressed in units of $\tau_{cr}$.}

% \clearpage
% 
%   %% For the preprint, redefine \figcaption to actually include 
%   %% the PS figure 
%   \newlength{\FigWidth}
%   \renewcommand{\figcaption}[2][FileNotFound.ps]{\begin{figure}
%       \includegraphics[width=\FigWidth]{#1}
%       \caption{#2}
%     \end{figure}}
%   \nonstopmode % set this in case the PS files can't be found
% 
% %\begin{center}
% %{\bf FIGURE CAPTIONS}
% %\end{center}
% 
% %\placefigure{fig1.ps}
% \setlength{\FigWidth}{\textwidth}
% \figcaption[fig1.ps]{Volume rendering (log scale) of the magnetic pressure
% for the $\beta=4$ simulation (strong field) at $t=2.25 \tau_{cr}$.
% Superimposed in the figure are the magnetic field lines.
% \label{f_lines}}
% 
% %\placefigure{fig2.ps}
% \setlength{\FigWidth}{0.75\textwidth}
% \figcaption[fig2.ps]{Volume rendering of the cloud density (log scale) for the
% $\beta=\infty$ simulation (left column), $\beta=100$ (center column), and
% $\beta=4$ (right column). 
% Time is expressed in units of $\tau_{cr}$.
% \label{den_hb}}

\end{document}